\documentclass[aps,prb,reprint]{revtex4-1}
\pdfoutput=1

\usepackage{amsmath}
\usepackage{amsfonts}
\usepackage{amssymb}
\usepackage{braket}
\usepackage{relsize}
\usepackage{hyperref}
\newcommand{\us}[1]{_{\textsc{\relsize{-2}{\textsl{#1}}}}}

\begin{document}


\title{Super-GCA from $\mathcal{N} = (2,2)$ Super-Virasoro}

\author{Ipsita Mandal}
\email{imandal@perimeterinstitute.ca}
\affiliation{Perimeter Institute for Theoretical Physics, Waterloo, Ontario N2L 2Y5, Canada}

\author{Ahmed Rayyan}
\email{arayyan@ualberta.ca}
\affiliation{University of Alberta, Edmonton, Alberta T6G 2R3, Canada}

\begin{abstract}
We derive the extended Supersymmetric Galilean Conformal Algebra (SGCA) in two spacetime dimensions by the method of group contraction on $2d$ $\mathcal{N}=(2,2)$ superconformal algebra. Both the parent and daughter algebras are infinite-dimensional. We provide the representation theory of the algebra. We adopt a superspace formalism for the SGCA fields, allowing us to write them down in a compact notation as components of superfields. We also discuss correlation functions, short supermultiplets and null states.
\end{abstract}

\maketitle


\section{Introduction}
\label{intro}

Recently there has been an extensive study of the Galiliean Conformal Algebra (GCA)~\cite{gca,ips2009,ba10,bagchi11,bagchi13,hagen,barnich,oblak,luk,luk2,duval,ayan,mart,rouhani,ali,arjuntensionless,arjunym} and its various supersymmetric extensions~\cite{sgca4d,ma10,aizawa0,gao,aizawa1,masterov12,masterov15,sakaguchi}. These algebras exhibit non-relativistic conformal symmetry and are obtained by a parametric contraction of the corresponding ``parent" relativistic conformal or superconformal group. In two dimensions, the parent Virasoro or super-Virasoro algebra itself is infinite-dimensional, and we can systematically obtain the $2d$ (S)GCA by contraction of the two copies of the (super-)Virasoro algebra.
$2d$ (S)CFTs are also important as they provide the necessary tool to formulate the worldsheet picture of the (super)string theories. It  has been shown~\cite{boucher,sen,banks} that superconformal algebras with $\mathcal{N} =2 $  and $\mathcal{N} =4 $  worldsheet  supersymmetry  describe  string compactification on complex manifolds of $SU(n)$ holonomy.

The $2d$ bosonic case was studied in great detail in~\cite{ba10}. The minimal supersymmetric extension giving us a supersymmetric GCA (SGCA) was obtained in~\cite{ma10} from the contraction of $\mathcal{N}=(1,1)$ super-Virasoro algebra~\cite{qiu84,qiu85,qiu86,berhadsky,goddard,sotkov}. In the present work, we will derive an extended SGCA from the parent SCFT with $\mathcal{N}=(2,2)$ supercharges~\cite{Zamolodchikov86,boucher,di85,di86,di86-2,nam,dobrev,matsuo,kiritsis,gi97,we90,blu09}. We will restrict to the Neveu-Schwarz sector so that we can use the superspace formalism. In any case, the $\mathcal{N} =2 $ Ramond and Neveu-Schwarz algebras are isomorphic by spectral flow~\cite{seiberg}. Such an SCFT could  prove useful in the  construction  of $2d$  critical systems with  a  hidden  $\mathcal{N} =2 $ supersymmetry. It has extra features like R-symmetry, chiral primaries and BPS bounds as compared to the $\mathcal{N}=(1,1)$ case. We will explore what new features emerge in the corresponding non-relativistic version.

The paper is organized as follows: In  Sec.~\ref{algebra}, we derive the $2d$ extended SGCA from the $(2,0)$ holomorphic and the $(0,2)$ anti-holomorphic super-Virasoro algebras by the method of group contraction~\cite{inonu}. In Sec.~\ref{rep}, we discuss the representation theory of the algebra. Sec.~\ref{superspace} provides the superspace formalism for the SGCA fields, allowing us to write them down in a compact notation as components of superfields. This section also discusses correlation functions, short supermultiplets and null states. Sec.~\ref{append} discusses the possibility of extending the R-symmetry of the non-relativistic algebra. We conclude with a summary and some outlook in Sec.~\ref{summary}.

\section{$2d$ SGCA from $ {\mathcal{N}} = (2,2)$ SCFT }
\label{algebra}

The $\mathcal{N} = 2$ super-Virasoro algebra is given by~\cite{blu09}:
\begin{equation}
\begin{split}
\label{sva}
& [\mathcal{L}_m, \mathcal{L}_n] =  (m-n) \mathcal{L}_{m+n} 
+ \frac{c}{8} \, m \, (m^2 - 1) \, \delta_{m+n,0} \, ,\\
& \left\{\mathcal{G}_r^+, \mathcal{G}_s^-\right\} =  \mathcal{L}_{r+s} 
+ \frac{1}{2}(r-s) \,J_{r+s} + \frac{c}{4}\left(r^2 - \frac{1}{4}\right)\delta_{r+s,0} \, ,\\
& [J_m, J_n] = \frac{c}{2} \, m \, \delta_{m+n,0} \, ,
\qquad [J_m, \mathcal{G}_r^\pm] = \pm \, \mathcal{G}_{m+r}^\pm \, ,\\
& [\mathcal{L}_m, \mathcal{G}_r^\pm]  = \left(\frac{m}{2} - r\right) \mathcal{G}_{m+r}^\pm \,,  
\qquad [\mathcal{L}_m, J_n] = -n\,J_{m+n} \, ,
\end{split}
\end{equation}
where $m, n \in \mathbb{Z}$ and $r, s \in \mathbb{Z}+ \lambda$. Furthermore, $\lambda = 0$ in the Ramond sector and $ \lambda  = \tfrac{1}{2}$ in the Neveu-Schwarz sector. Only the Neveu-Schwarz sector will be considered in this paper and hence $r, s \in \mathbb{Z} + \tfrac{1}{2}$ in all subsequent discussions.

Dealing with extended supersymmetry implies the presence of an R-symmetry. Here, this is represented by the bosonic $U(1)$ current algebra generated by $J_n$. Hence, in contrast to the $\mathcal{N}  =  0,  1$  cases, the  $\mathcal{N}  =2$  superconformal  algebra  has  highest-weight  states  represented  by  two  parameters,  usually denoted by $h$ and $q$,  corresponding to the eigenvalues of the two elements ($\mathcal{L}_0$ and $J_0$) of the Cartan  subalgebra. The $\mathcal{G}_r^\pm$ are fermionic operators with charges $\pm1$ with respect to the U(1) current.


In general, a superconformal field theory (SCFT) in $2d$ comes with a holomorphic as well as an anti-holomorphic copy of~\eqref{sva}. The two copies are identical in structure, and one can obtain the anti-holomorphic copy by replacing $\mathcal{L}_n$ with $\bar{\mathcal{L}}_n$, $\mathcal{G}_r^\pm$ with $\bar{\mathcal{G}}_r^\pm$, $J_n$ with $\bar{J}_n$, and $c$ with $\bar{c}$. Note that the independence of the two algebras implies that the (anti-)commutator of an unbarred operator with a barred one vanishes. 

Now, a contraction of both copies of~\eqref{sva} is performed by defining the following generators:
\begin{equation}
\begin{split}
\label{sgcadef}
L_n &= \lim_{\epsilon \to 0} \,
(\bar{\mathcal{L}}_n + \mathcal{L}_n)\, , 
 \quad M_n = \lim_{\epsilon \to 0} \, \epsilon \, (\bar{\mathcal{L}}_n - \mathcal{L}_n)\, ,\\
G_r^\pm &= \lim_{\epsilon \to 0} \, (\bar{\mathcal{G}}_r^\pm + \mathcal{G}_r^\pm) \, ,  
\quad H_r^\pm = \lim_{\epsilon \to 0} \,  \epsilon  \, (\bar{\mathcal{G}}_r^\pm - \mathcal{G}_r^\pm)\, ,\\
I_n &= \lim_{\epsilon \to 0} \, (\bar{J}_n + J_n)\, ,
\quad S_n = \lim_{\epsilon \to 0} \epsilon   \, (\bar{J}_n - J_n)\, .\\
\end{split}
\end{equation}
The scaling chosen corresponds to a non-relativistic scaling of coordinates so that the velocities $v \sim \epsilon$ (see~\cite{gca,ba10,ma10} for more details).

According to~\eqref{sva}, the non-zero (anti-)commutators of the generators are:
\begin{equation}
\begin{split}
\label{sgca}
& [L_m, L_n] =  (m-n)\, L_{m+n} + C_1 \, m\,(m^2 - 1)\, \delta_{m+n,0}\, ,\\
& [L_m, M_n]  =  (m-n) \,M_{m+n} + C_2 \, m\,(m^2 - 1) \, \delta_{m+n,0}\,,\\
& [I_m, I_n]  = 4\, C_1 \, m\, \delta_{m+n,0} \,, 
\quad [I_m, S_n] = 4 \, C_2 \, m \, \delta_{m+n,0}\, ,\\
& [L_m, I_n] = -n  \, I_{m+n}\,,\quad
[L_m, S_n] = -n   \, S_{m+n} = [M_m, I_n]\,,\\
& \left\{G_r^+, G_s^-\right\} =  L_{r+s} + \frac{1}{2}(r-s)I_{r+s}  
+ 2  \, C_1\left(r^2 - \frac{1}{4}\right)  \, \delta_{r+s,0}\, ,\\
&  \left\{G_r^+, H_s^-\right\} =  \left\{H_r^+, G_s^-\right\} \,\\
& = M_{r+s} + \frac{1}{2}(r-s)   S_{r+s} 
+ 2  \, C_2\left(r^2 - \frac{1}{4}\right)\delta_{r+s,0} \ ,\\
& [L_m, G_r^\pm] = \left(\frac{m}{2} - r\right) G_{m+r}^\pm \, ,\\
& [L_m, H_r^\pm]  = [M_m, G_r^\pm] = \left(\frac{m}{2} - r\right) H_{m+r}^\pm \,,\\
& [I_m, G_r^\pm]  = \pm  \,  G_{m+r}^\pm \, , \quad
[I_m, H_r^\pm] = [S_m, G_r^\pm] =  \pm   \, H_{m+r}^\pm \,,\\
\end{split}
\end{equation}
where the central charges are given by:
\begin{equation}
\label{ccharge}
C_1 =\lim_{\epsilon \to 0} \frac{ \bar c + c} {8}
 \,, \quad  C_2 = \lim_{\epsilon \to 0} \epsilon \,\frac{ \bar c - c} {8}\,.
\end{equation}
This is the $2d $ supersymmetric Galilean conformal algebra (SGCA) with four supercharges.

\section{Representation Theory of the Extended SGCA}
\label{rep}
From the (anti-)commutation relations~\eqref{sgca}, one can check that the Cartan subalgebra of our SGCA is generated by $L_0$, $M_0$, $I_0$ and $S_0$. Hence it will be convenient to construct representations by considering states having definite weights which include the scaling dimension $\Delta$, the `rapidity' $\tilde \Delta$, and the `charges' $\kappa$ and $\tilde \kappa$, such that:
\begin{equation}
\begin{split}
& L_0 \ket{\Delta, \tilde{\Delta}, \kappa, \tilde{\kappa}} = \Delta \ket{\Delta, \tilde{\Delta}, \kappa, \tilde{\kappa}}  , \\
& I_0 \ket{\Delta, \tilde{\Delta}, \kappa, \tilde{\kappa}} = \kappa \ket{\Delta, \tilde{\Delta}, \kappa, \tilde{\kappa}} ,  \\
& M_0 \ket{\Delta, \tilde{\Delta}, \kappa, \tilde{\kappa}}  = \tilde{\Delta}  \ket{\Delta, \tilde{\Delta}, \kappa, \tilde{\kappa}}  , \\
& S_0\ket{\Delta, \tilde{\Delta}, \kappa, \tilde{\kappa}}  = \tilde{\kappa}\ket{\Delta, \tilde{\Delta}, \kappa, \tilde{\kappa}} .
\nonumber  \\
\end{split}
\end{equation}
We can relate these weights to the weights $h$, $\bar{h}$, $q$ and $\bar{q}$ in the parent SCFT, which are the eigenvalues of $\mathcal{L}_0 $, $\bar{\mathcal{L}}_0 $,  $J_0 $ and $\bar{J}_0 $ respectively. The relations follow from the definition of our SGCA operators in~\eqref{sgcadef}:
\begin{equation}
\begin{split}
\label{weight}
\Delta &= \lim_{\epsilon \to 0} \, (\bar{h} + h)\, , 
 \quad \tilde{\Delta} = \lim_{\epsilon \to 0} \, \epsilon  \, (\bar{h} - h) \, ,\\
\kappa &= \lim_{\epsilon \to 0} \, (\bar{q} + q) \, , 
\quad \tilde{\kappa} = \lim_{\epsilon \to 0}  \, \epsilon  \, (\bar{q} - q) \, .\\
\end{split}
\end{equation}

Using~\eqref{sgca}, one can investigate the action of the SGCA generators on these weights. For $\Delta$, one finds that, for an arbitrary SGCA generator labelled generically as $\mathcal{W}_{\nu} $ (with $ \nu  \in \mathbb{Z}$ for the bosonic generators and $\nu  \in \mathbb{Z} + \frac{1}{2}$ for the fermionic generators),
\begin{equation}
L_0 \, \mathcal{W}_\nu \ket{\Delta, \tilde{\Delta}, \kappa, \tilde{\kappa}} = (\Delta - \nu )\ket{\Delta, \tilde{\Delta}, \kappa, \tilde{\kappa}} . 
\end{equation}

So generators with $ \nu  > 0$ lower the value of $\Delta$, while generators with $ \nu < 0$ raise it.
For $\kappa$, one finds that a bosonic $\mathcal{W}_\nu $ leaves it unchanged, whereas:
\begin{equation}
\begin{split}
& I_0 \, G_r^\pm \ket{\Delta, \tilde{\Delta}, \kappa, \tilde{\kappa}} = (\kappa \pm 1)\, G_r^\pm \ket{\Delta, \tilde{\Delta}, \kappa, \tilde{\kappa}} , \\
 & I_0 \,H_r^\pm \ket{\Delta, \tilde{\Delta}, \kappa, \tilde{\kappa}} = (\kappa \pm 1)H_r^\pm \ket{\Delta, \tilde{\Delta}, \kappa, \tilde{\kappa}}  .
 \end{split}
\end{equation}

Demanding that the weights be bounded from below implies the existence of primary states $\ket{p} \equiv \ket{\Delta, \tilde{\Delta}, \kappa, \tilde{\kappa} }_p$, defined by:
\begin{equation}
\begin{alignedat}{3}
&L_n \ket{p} = M_n \ket{p} = I_n \ket{p} = S_n \ket{p} = 0 \quad &(\, \forall \,  n > 0 \, )\, ,\\
&G_r^\pm \ket{p} = H_r^\pm \ket{p} = 0 \quad &( \, \forall \, r > 0 \, ) \, .
\end{alignedat}
\end{equation}

As a result, given a primary state $\ket{p}$, one can get descendent states by the repeated action of $L_{-n}$, $M_{-n}$, $G_{-r}^\pm$, $H_{-r}^\pm$, $I_{-n}$, and $S_{-n}$ (with $n$, $r$ $> 0$) on it. The primary state with all its possible descendant states form a representation of the SGCA.

One interesting note to make is that $I_0$ commutes with all $L_{-n}, M_{-n}, I_{-n}$ and $S_{-n}$. As a result, descendant states formed by the action of only those four operators will still be eigenstates of $I_0$, and so they all share the same $\kappa$. Similarly, the action of $M_{-n}, H_{-n}^\pm$, and $S_{-n}$ on a primary state yields eigenstates of $M_0$ with the same value of $\tilde{\Delta}$. Finally, the action of $L_{-n}, M_{-n}, H_{-r}^\pm, I_{-n}$ and $S_{-n}$ on a primary state yields eigenstates of $S_0$ with the same value of $\tilde{\kappa}$.

We require that the vacuum state $\ket{0}$ be invariant under the action of the globally defined sector of the SGCA. This corresponds to $\ket{0}$ satisfying the following properties:
\begin{equation}
\begin{alignedat}{3}
\label{vacuum}
L_n\ket{0} &= M_n\ket{0} = 0\quad ( \mbox{ for }  n \geq -1 \, )\, ,\\
I_n\ket{0} &= S_n\ket{0} = 0 \quad( \mbox{ for } n \geq 0 \, ) \, ,\\
G_r^\pm\ket{0} &= H_r^\pm\ket{0} = 0 \quad( \mbox{ for } r \geq -\tfrac{1}{2} \, )\, .
\end{alignedat}
\end{equation}

\section{Extended SGCA on Superspace}
\label{superspace}

We will use the superspace formalism of the $2d$ $\mathcal{N} = (2,2)$ SCFT~\cite{we90,kiritsis} for the Neveu-Schwarz sector. The extended supersymmetry requires the enlargement of the space to include two ``fermionic" coordinates in each of the two sectors. Specifically, the $(2,0)$ holomorphic and the $(0,2)$ anti-holomorphic sectors are represented by the coordinates
\begin{equation}
\mathcal{Z} \equiv (z, \theta^+, \theta^-)\quad \text{and} \quad \mathcal{\bar{Z}} \equiv (\bar{z}, \bar{\theta}^+, \bar{\theta}^-)
\end{equation}
respectively. For the remainder of this section, we will only consider the holomorphic sector.

A superfield is a function at most linear in each of its Grassmann variables, due to their inherent nilpotency. As such, any superfield in the $(2,0)$ superspace can be expanded as
\begin{equation}
F(\mathcal{Z}) = f(z) + \theta^+ \psi^-(z) + \theta^- \psi^+(z) + \theta^+\theta^-g(z) \, ,
\end{equation}
where $f(z)$ and $g(z)$ are bosonic fields, while $\psi^-(z)$ and $\psi^+(z)$ are fermionic fields. 
A primary superfield is defined as a superfield generating a highest-weight irreducible representation of the
$(2,0)$ superconformal algebra.

The generators of the superanalytic transformations in the superspace can be represented as~\cite{we90}:
\begin{equation}
\begin{split}
\label{transform}
\mathcal{L}_n & \equiv  
z^{ n+1 } \partial_z + \frac{n+1} {2} z^n 
\left ( \theta^+ \partial_{\theta^+} + \theta^- \partial_{\theta^-}\right )  , \\
J_n & \equiv z^n 
 \left (  \theta^- \partial_{\theta^-}   -   \theta^+ \partial_{\theta^+}  \right ),\\
\frac {\mathcal{G}_r^+ }     {\sqrt 2 }& \equiv 
z^{r + \frac{1}{2} }
\left ( \partial_{\theta^+}  - \frac{1}{2} \theta^- \partial_{z} \right ) 
+ \frac{1}{2}  \left(  r + \frac{1}{2} \right)  z^{r -\frac{1}{2}}  \theta^+ \theta^- \partial_{\theta^+ } \, , \\
\frac{\mathcal{G}_r^- } {\sqrt 2 }  & \equiv 
z^{r + \frac{1}{2} }
\left ( \partial_{\theta^-}  - \frac{1}{2} \theta^+ \partial_{z} \right ) 
- \frac{1}{2} \left(  r + \frac{1}{2} \right)  z^{r -\frac{1}{2} }  \theta^+ \theta^- \partial_{\theta^- } \, .
\end{split}
\end{equation}
The corresponding differential operators acting on a primary superfield $  \mathcal {F}  (\mathcal{Z})$ are given as:
\begin{widetext}
\begin{equation}
\begin{split}
\label{svadiff}
[\mathcal{L}_n, \mathcal {F} ] &= z^n\left[z \,  \partial_z + \left(\frac{n+1}{2}\right) (\theta^+ \partial_{\theta^+} + \theta^- \partial_{\theta^-}) + (n+1)\left(h + \frac{n\, q}{4}z^{-1} \theta^+\theta^-\right)\right]  \mathcal {F}  ,\\
[\mathcal{G}_r^\pm,  \mathcal {F}  ] &= 
\sqrt 2   \left[z^{r+ \frac{1}{2} }\left(\partial_{\theta^\pm} - \frac{1}{2}\theta^\mp\partial_z\right) - \left(r+\frac{1}{2}\right)z^{r-\frac{1}{2}}\theta^\mp\left(\frac{1}{2}\theta^\pm\partial_{\theta^\pm} +h \mp \frac{q}{2}\right)\right]  \mathcal {F}  ,\\
[J_n,  \mathcal {F}  ] &= z^n\left[\theta^- \partial_{\theta^-} - \theta^+ \partial_{\theta^+} 
+ n\, h \, z^{-1}\theta^+\theta^- + q \, \right]  \mathcal {F} .
\end{split}
\end{equation}
\end{widetext}
The transformations for the anti-holomorphic sector take an identical form, with $z$ and $\theta^\pm$ replaced by $\bar{z}$ and $\bar{\theta}^\pm$, respectively.

Now we construct the superspace formalism for the SGCA by taking the non-relativistic limit of these superspace coordinates. Our new coordinates are obtained by taking the linear combinations 
\begin{equation}
\begin{split}
 t = \frac{ z+\bar{z} } {2} \, , \, \, x = \frac{z-\bar{z}}{2} \, ,\,\,
 \alpha^\pm = \frac{ \theta^\pm+\bar{\theta}^\pm} {2}\, , 
\, \, \beta^\pm =  \frac{\theta^\pm-\bar{\theta}^\pm } {2} \, ,
\end{split}
\end{equation}
and then taking the scalings as:
\begin{equation}
\label{scale}
t \to t \, , \quad x \to \epsilon \, x\, ,
 \quad  \alpha^\pm \to \alpha^\pm, 
 \quad \beta^\pm \to \epsilon \, \beta^\pm .
\end{equation}
Hence an extended SGCA primary superfield is of the form
\begin{widetext}
\begin{equation}
\begin{split}
\label{superfield}
 \Phi(t,x, \alpha^\pm, \beta^\pm) 
 & = p (t,x) + \alpha^+ \gamma(t,x) + \beta^+ \bar{\gamma}(t,x) 
 + \alpha^+ \beta^+ d(t,x)
 +\alpha^-[ \, \delta(t,x) + \alpha^+ e(t,x) + \beta^+ f(t,x) + \alpha^+ \beta^+\bar{\delta}(t,x)\, ]
 \\
& \, \, \,+\beta^-[ \, \eta(t,x) + \alpha^+ g(t,x) + \beta^+ \ell (t,x) + \alpha^+ \beta^+\bar{\eta}(t,x)\, ]\\
&  \, \, \, +\alpha^-\beta^-[ \,  j(t,x) + \alpha^+ \zeta(t,x) + \beta^+ \bar{\zeta}(t,x) + \alpha^+ \beta^+ s(t,x)\, ] \, ,
\end{split}
\end{equation}
\end{widetext}
where $p(t,x)$ is a primary field with respect to the bosonic generators. The bosonic and the fermionic fields have been denoted by Latin and Greek characters respectively.
The group contraction in~\eqref{sgcadef} implies that the SGCA operators should act on a primary superfield $\Phi$ as:
\begin{equation}
\begin{split}
[L_n, \Phi] &= \lim_{\epsilon \to 0} \, [\bar{\mathcal{L}}_n + \mathcal{L}_n,  \mathcal {F}]\,,  
\quad [M_n, \Phi] = \lim_{\epsilon \to 0} 
\epsilon \, [\bar{\mathcal{L}}_n - \mathcal{L}_n,  \mathcal {F}] \,,\\
[G_r^\pm, \Phi] &= \lim_{\epsilon \to 0}\, [\bar{\mathcal{G}}_r^\pm + \mathcal{G}_r^\pm,  \mathcal {F}]\,,  
\quad [H_r^\pm, \Phi] = \lim_{\epsilon \to 0} \, \epsilon\, [\bar{\mathcal{G}}_r^\pm - \mathcal{G}_r^\pm,  \mathcal {F}] \, ,\\
[I_n,\Phi] &= \lim_{\epsilon \to 0} \, [\bar{J}_n + J_n,  \mathcal {F} ]\,, 
\quad [S_n,\Phi] = \lim_{\epsilon \to 0} \, \epsilon \, [\bar{J}_n - J_n,  \mathcal {F} ] \, .\\
\end{split}
\end{equation}

Taking the scaling limit~\eqref{scale} with~\eqref{svadiff} and its anti-holomorphic counterpart, one finally arrives at
\begin{widetext}
\begin{equation}
\begin{split}
[L_n, \Phi]  &= t^n
 \bigg\{
t \, \partial_t + (n+1)x \, \partial_x + \left(\frac{n+1}{2}\right)(\alpha^+\partial_{\alpha^+} + \beta^+\partial_{\beta^+} + \alpha^-\partial_{\alpha^-} + \beta^-\partial_{\beta^-}) 
+\frac{n \, (n+1)}{2} \frac{x} {t}(\alpha^+\partial_{\beta^+} + \alpha^-\partial_{\beta^-} ) \\
&\quad +(n+1)\bigg( \Delta + \frac{n \, \kappa}{4 \, t} \alpha^+\alpha^-
- \frac{n} {t}
\left[  \tilde{\Delta} \, x + \frac{\tilde{\kappa}}{4}
\left ( \, \frac{x  (n-1)} {t}\alpha^+\alpha^- + \alpha^+\beta^- + \beta^+\alpha^- \right )\right]\bigg)
\bigg\} \,  \Phi \, , \\
[M_n, \Phi] &= t^n
\left\{-t \, \partial_x - \left(\frac{n+1}{2}\right)(\alpha^+\partial_{\beta^+} + \alpha^-\partial_{\beta^-}) + (n+1)\left(\tilde{\Delta} + \frac{n\, \tilde{\kappa}}{4 \, t} \alpha^+\alpha^-\right)\right\}  \, \Phi \, ,\\
[G_r^\pm, \Phi] 
&= \sqrt 2  \,  \bigg\{ t^{r+\frac{1}{2}}
\left[ \left(\partial_{\alpha^\pm} - \frac{1}{2}(\alpha^\mp\partial_t + \beta^\mp\partial_x\right)+\left(r+\frac{1}{2}\right) \frac{x} {t} \left(\partial_{\beta^\pm} - \frac{1}{2} \alpha^\mp \partial_x\right)  \right] \\
& \quad +\left(r+\frac{1}{2}\right)t^{r-\frac{1}{2}}\bigg[\left(\beta^\mp +\Big(r-\frac{1}{2}\Big)
\frac{x} {t} \,  \alpha^\mp\right)\left(\tilde{\Delta} \mp \frac{\tilde{\kappa}}{2} - \frac{1}{2}\alpha^\pm\partial_{\beta^\pm}\right) 
 -\alpha^\mp\Big(\Delta \mp \frac{\kappa}{2} + \frac{1}{2}(\alpha^\pm\partial_{\alpha^\pm} + \beta^\pm\partial_{\beta^\pm})\Big)\bigg]\bigg\} \, \Phi \,, \\
[H_r^{\pm} , \Phi] &=
 \sqrt 2  \,   \left\{ t^{r+\frac{1}{2}} \left[\frac{1}{2} \alpha^\mp\partial_x - \partial_{\beta^{\pm}}\right]
+ \left(r+\frac{1}{2} \right)
t^{r-\frac{1}{2}}  \, \alpha^\mp     \left[\frac{1}{2}\alpha^\pm\partial_{\beta^\pm} \pm 
\Big(\frac{\tilde{\kappa}}{2} \mp \tilde{\Delta}\Big)\right]    \right\} \, \Phi   \, , \\
[I_n, \Phi] &=  t^n
\Big\{\alpha^-\partial_{\alpha^-} + \beta^-\partial_{\beta^-} - \alpha^+\partial_{\alpha^+} - \beta^+\partial_{\beta^+} + n\,\frac{x} {t}(\alpha^-\partial_{\beta^-} - \alpha^+\partial_{\beta^+}) \\
& \quad - \frac{n} {t} \left[\tilde{\Delta}\big(\alpha^+\beta^- + \beta^+\alpha^- + (n-1)\, \frac{x} {t} \, \alpha^+\alpha^-\big) + \tilde{\kappa}\, x \right]
+ \frac{n} {t} \, \Delta \,  \alpha^+\alpha^- + \kappa \Big\} \, \Phi   \, , \\
[S_n, \Phi] &=    t^n \left\{\alpha^+\partial_{\beta^+} - \alpha^-\partial_{\beta^-} 
+ \frac{n} {t} \tilde{\Delta} \, \alpha^+\alpha^- + \tilde{\kappa}\right\}   \, \Phi \, .
\label{sgcadiff}
\end{split}
\end{equation}
\end{widetext}

Assuming that the operator-state correspondence present in the parent $2d$ SCFT continues to hold in the SGCA $\left ( \mbox{i.e.\ } \mathcal{O }(t,x)  \leftrightarrow  \mathcal{ O }(0,0)\ket{0}  \right )$ , we find that
the primary state $\ket{p} \equiv p(0,0)\ket{0}$ transforms as:
\begin{equation}
\begin{split}
& G_{-\frac{1}{2}}^+ \ket{p}  =\sqrt 2  \ket{\gamma},\; 
G_{-\frac{1}{2}}^-\ket{p} = \sqrt 2  \ket{\delta},\nonumber
\end{split}
 \end{equation}
 \begin{equation}
\begin{split}
& H_{-\frac{1}{2}}^+  \ket{p} = -\sqrt 2 \ket{\bar{\gamma}},\;
 H_{-\frac{1}{2}}^-\ket{p}  = -  \sqrt 2 \ket{\eta},\nonumber
 \end{split}
 \end{equation}
\begin{equation}
\begin{split}
& G_{-\frac{1}{2}}^+     G_{-\frac{1}{2}}^-\ket{p}  =   L_{-1}\ket{p} - 2  \ket{e},\;
 G_{-\frac{1}{2}}^+H_{-\frac{1}{2}}^+\ket{p} = -2  \ket{d}  ,\nonumber
\end{split}
 \end{equation}
 \begin{equation}
\begin{split}
&
 G_{-\frac{1}{2}}^-H_{-\frac{1}{2}}^+\ket{p} =    M_{-1}\ket{p} - 2 \ket{f},\;
 G_{-\frac{1}{2}}^-  H_{-\frac{1}{2}}^-\ket{p} = - 2 \ket{j},\nonumber
\end{split}
 \end{equation}
 \begin{equation}
\begin{split}
& 
G_{-\frac{1}{2}}^+ H_{-\frac{1}{2}}^-\ket{p} =   M_{-1} \ket{p} + 2  \ket{g},\;
H_{-\frac{1}{2}}^+ H_{-\frac{1}{2}}^-\ket{p} =  - 2 \ket{\ell},\nonumber
\end{split}
 \end{equation}
 \begin{equation}
\begin{split}
&
 G_{-\frac{1}{2}}^+  G_{-\frac{1}{2}}^-    H_{-\frac{1}{2}}^+\ket{p}  = 
\sqrt 2
 \left ( M_{-1}\ket{\gamma} - L_{-1}\ket{\bar{\gamma}} + 2  \ket{\bar{\delta}}   \right ) ,
\nonumber
\end{split}
 \end{equation}
 \begin{equation}
\begin{split}
&
 G_{-\frac{1}{2}}^+ H_{-\frac{1}{2}}^+  H_{-\frac{1}{2}}^-\ket{p}  =\sqrt 2
 \left (  M_{-1}\ket{\bar{\gamma}} + 2 \ket{\bar{\eta}} \right ) ,
\nonumber
\end{split}
 \end{equation}
 \begin{equation}
\begin{split}
&
 G_{-\frac{1}{2}}^-H_{-\frac{1}{2}}^+H_{-\frac{1}{2}}^-\ket{p} 
 = -  \sqrt 2
 \left (  M_{-1}\ket{\eta} + 2  \ket{\bar{\zeta}} \right ) ,
\nonumber
\end{split}
 \end{equation}
 \begin{equation}
\begin{split}
& H_{-\frac{1}{2}}^-G_{-\frac{1}{2}}^+G_{-\frac{1}{2}}^-\ket{p} = 
\sqrt 2   \left (  L_{-1}\ket{\eta} - M_{-1}\ket{\delta}  - 2  \ket{\zeta} \right )   ,
\nonumber
\end{split}
 \end{equation}
 \begin{equation}
\begin{split}
&
 G_{-\frac{1}{2}}^+G_{-\frac{1}{2}}^-H_{-\frac{1}{2}}^+H_{-\frac{1}{2}}^-\ket{p} 
= 
 (M_{-1})^2 \ket{p} 
+ 2   M_{-1} \left ( \ket{g} - \ket{j}  \right ) \\
& \qquad \qquad 
\quad \quad \quad \qquad \quad \,\,\, + 4  L_{-1}\ket{\ell} -4 \ket{s} .
\end{split}
\end{equation}
In essence, given a primary superfield, we can jump around the components of the superfield by these operations. This is expected as the primary superfields comprise the irreducible representations of the SGCA.

\subsection{Correlation Functions}
\label{corrln}

We now construct correlation functions obeying the SGCA invariance. One method is to directly use the commutators of the SGCA, but the mixing of holomorphic and anti-holomorphic algebras leads to very complicated expressions for the differential operators. Another method is to construct the correlation functions respecting the super-Virasoro algebra, and then taking a non-relativistic scaling of the coordinates and weights to obtain the SGCA result. We will demonstrate the latter method for simplicity.

The two-point function of the $(2,0)$ super-Virasoro algebra is given by~\cite{we90}:
\begin{equation}
\begin{split}
&  G^{(2) } \us{SVA}(z\us{12}, \theta^\pm\us{12}) \equiv 
\braket{\mathcal{F} \us{1}(x\us{1}, t\us{1}, \alpha^\pm\us{1}, \beta^\pm\us{1})  \, 
\mathcal{F} \us{2}(x\us{2}, t\us{2}, \alpha^\pm\us{2}, \beta^\pm\us{2})}   \\
&    =
\delta_{ h_1 , h_2 }     \, \delta_{q_1, -q_2}  \,
z\us{12}^{-2h\us{1}} \left(1-\frac{q\us{1}}{2} z\us{12}^{-1} \theta\us{12}^+ \theta\us{12}^-\right) ,
\end{split}
\end{equation}
where  $z\us{12} = z\us{1} - z\us{2} - \frac{1}{2}(\theta^+\us{1}\theta^-\us{2} + \theta^-\us{1} \theta^+\us{2})$
and $\theta^\pm\us{12} = \theta^\pm\us{1} - \theta^\pm\us{2}$. Also the overall multiplicative constant has been set to unity by adjusting the normalization of the operators.
The form of the superintervals are fixed by the invariance under the operators $\mathcal{L}\us{-1}$ and $\mathcal{G}_{-\frac{1}{2}}^\pm$. Invariance under $\mathcal{J}\us{0}$ and $\mathcal{L}\us{1}$ requires that the charges of the two superfields are related as $q\us{1} = -q\us{2},\; h\us{1} = h\us{2}$ for a non-zero answer. Similarly, one can find a $\bar{G}\us{SVA}$ corresponding to the anti-holomorphic copy of the algebra, with all the intervals and the weights replaced with their barred counterparts.

Using the scaling in \eqref{scale}, one obtains the following SGCA superintervals:
\begin{equation}
\begin{split}
& x\us{12} = x\us{1} - x\us{2} - \frac{1}{2}(\alpha^+\us{1}\beta^-\us{2} + \alpha^-\us{1} \beta^+\us{2} + \beta^+\us{1}\alpha^-\us{2} + \beta^-\us{1} \alpha^+\us{2}) , \\
 & t\us{12} = t\us{1} - t\us{2} - \frac{1}{2}(\alpha^+\us{1}\alpha^-\us{2} + \alpha^-\us{1} \alpha^+\us{2}),\\
 & \alpha^\pm\us{12} = \alpha^\pm\us{1} - \alpha^\pm\us{2} ,\;
\beta^\pm\us{12} = \beta^\pm\us{1} - \beta^\pm\us{2} .
\end{split}
\end{equation}
Now $G^{(2) } \us{SCFT} \equiv  G^{(2) } \us{SVA} \, \bar{G}^{(2) } \us{SVA}$ can be scaled to obtain our desired result:
\begin{widetext}
\begin{eqnarray}
&& G^{(2) } \us{SGCA}(x\us{12}, t\us{12}, \alpha^\pm\us{12},\beta^\pm\us{12})  \equiv 
\braket{\Phi\us{1}(x\us{1}, t\us{1}, \alpha^\pm\us{1}, \beta^\pm\us{1})
\, \Phi\us{2}(x\us{2}, t\us{2}, \alpha^\pm\us{2}, \beta^\pm\us{2})} \nonumber \\
&& =
\delta_{ \Delta\us{1} , \Delta\us{2} }     \, \delta_{\tilde{\Delta}\us{1} , \tilde{\Delta}\us{2} } \,
\delta_{ \kappa\us{1} , - \kappa\us{2} }  \,
\delta_{  \tilde{\kappa}\us{1} ,  -\tilde{\kappa}\us{2} } \,
 \frac{  \exp \left ( \frac{ 2   \, \tilde{\Delta} \us{1}  x\us{12}}{t\us{12}} \right ) }  
 { t\us{12}^{    2   \Delta\us{1}} }  
 \left(1-\frac{\kappa\us{1}}{2} \frac{\alpha\us{12}^+ \alpha\us{12}^-}{t\us{12}}
+ \dfrac{\tilde{\kappa}\us{1}} {2}
 \frac{\alpha\us{12}^+\beta\us{12}^- 
+ \beta\us{12}^+\alpha\us{12}^- - \frac{x\us{12}}{t\us{12}}\alpha\us{12}^+\alpha\us{12}^-}{t\us{12}} - \dfrac{\tilde{\kappa}\us{1}^2} {4}\frac{ \alpha\us{12}^+\alpha\us{12}^-\beta\us{12}^+\beta\us{12}^-}{t\us{12}^2}  \right) .\nonumber \\
\end{eqnarray}
\end{widetext}
One can find the correlation functions of the component fields of $ \Phi\us{1} $ and $ \Phi\us{2} $ by expanding both sides in terms of the fermionic coordinates and equating the coefficients.


\subsection{Short Supermultiplets}
\label{bps}

Let us consider primaries satisfying $G_{-\frac{1}{2} }^{\mathcal{S}}  \ket{\Delta, \tilde{\Delta}, \kappa, \tilde{\kappa}}_p =0 $, where ${\mathcal{S}}$ takes the value $ + \mbox{ or } -  $. The anti-commutators
\begin{equation}
\begin{split}
\label{short1}
\left\{G_{-\frac{1}{2}}^+, G_{ \frac{1}{2}}^-\right\} &=  L_{ 0 } - \frac{ I_{ 0 } }{2},\quad
\left\{G_{-\frac{1}{2}}^+, H_{ \frac{1}{2}}^-\right\} =  M_{ 0 } - \frac{ S_{ 0 } } {2}  , \\
\left\{G_{-\frac{1}{2}}^-, G_{ \frac{1}{2}}^+ \right\} &=  L_{ 0 } + \frac{ I_{ 0 } }{2},\quad
\left\{G_{-\frac{1}{2}}^-, H_{ \frac{1}{2}}^+\right\} =  M_{ 0 } + \frac{ S_{ 0 } } {2} ,
\end{split}
\end{equation}
tell us that $ \Delta =\mathcal{S}  \kappa/ 2 $ and $ \tilde{\Delta} =\mathcal{S}  \tilde{\kappa} / 2  $ for these primaries. From~\eqref{weight}, we find that these conditions correspond to a chiral or anti-chiral primary of the parent SCFT depending on whether $\mathcal{S} $ takes the value $+$ or $-$. In other words, the parent primary satisfies $h = \mathcal{S} q /2$  and $\bar h = \mathcal{S} \bar q /2$ giving rise to BPS multiplets in both the holomorphic and anti-holomorphic sectors which are shorter than the generic multiplets. In the SGCA too, the above conditions lead to shortening of the generic number of components for a superfield.

If we consider primaries satisfying $H_{-\frac{1}{2} }^{\mathcal{S}}  \ket{\Delta, \tilde{\Delta}, \kappa, \tilde{\kappa}}_p =0 $ (where ${\mathcal{S}}$ can be $+$ or $-  $), the anti-commutators
\begin{equation}
\begin{split}
\label{short2}
& \left\{ H_{-\frac{1}{2}}^{\pm}, H_{ \frac{1}{2}}^{\pm} \right\} =0 , \quad 
\left\{ H_{-\frac{1}{2}}^+, G_{ \frac{1}{2}}^- \right\} =  M_{ 0 } - \frac{ S_{ 0 } } {2}  , \\
& \left\{ H_{-\frac{1}{2}}^-, G_{ \frac{1}{2}}^+ \right\} =  M_{ 0 } +  \frac{ S_{ 0 } } {2}  
\end{split}
\end{equation}
translate into $ \tilde{\Delta} =\mathcal{S}  \tilde{\kappa} / 2  $. These conditions also lead to short multiplets for the SGCA. However, the parent SCFT primaries need not be chiral or anti-chiral in this case. But if they are (anti-)chiral in the holomorphic sector, then they are forced to be (anti-)chiral also in the anti-holomorphic sector, and vice versa.

%

\subsection{Null States and Kac-like formula}

The null states obtained from the Kac-like formula~\cite{boucher,kiritsis} in the $\mathcal{N} = (2,2)$ SCFT will also give null states in the daughter SGCA once we translate the relations  involving SCFT weights into equations involving SGCA weights by taking the appropriate scaling limits. However, the relation~\eqref{ccharge} tells us that the parent SCFTs need not be unitary. So naturally, the number of SGCA null states will be larger than those obtained by the scaling method involving unitary SCFTs. The derivation of a generic formula will be a very difficult task and is left for future investigation. One can of course find the null states level by level from the intrinsic SGCA analysis. However, this procedure becomes more and more algebraically cumbersome as the level increases.
One can easily see that the condition on the weights of a primary state to get the lowest level null state (at level $\frac{1} {2}$) will actually coincide with the expressions derived for the short multiplets. One can further work out the Kac-like formula and fusion rules for the SGCA primaries. This is left for future work.

\section{Possibility of extending the R-symmetry}

\label{append}

Let us try to see if the SGCA with four supersymmetries can have an extended R-symmetry, other than the $U(1) \times U (1)$ that we have obtained by the group contraction. 

The first thing we can try is to examine whether we can promote the $U(1)$ current algebra generated by $ I_n$'s to an $ SU(2)$ current algebra generated by $ {\mathcal{J}}^j_n$'s such that we have the modified (anti-)commutators:
\begin{equation}
\begin{split}
\label{modi1}
 \left\{G_r^+, G_s^-\right\}  = & L_{r+s} + \frac{1}{2}(r-s) \, \sigma_{+-}^j \, {\mathcal{J}}^j_{r+s}  \\
& + 2  \, C_1\left(r^2 - \frac{1}{4}\right)  \, \delta_{r+s,0}\, ,\\
 \left\{G_r^-, G_s^+ \right\}  = & L_{r+s} + \frac{1}{2}(s-r) \, \sigma_{-+}^j \,  {\mathcal{J}}^j_{r+s}  \\
& + 2  \, C_1\left( s^2 - \frac{1}{4}\right)  \, \delta_{r+s,0}\, ,\\
\left\{ G_r^{ + }, G_s^{ + } \right\} =&  \frac{r-s}{2}  \,  {\mathcal{J}}^3_{r+s}\, ,\quad
\left\{ G_r^{ -  }, G_s^{ - } \right\} = - \frac{ r-s}{2}  \,\,  {\mathcal{J}}^3_{r+s} \, ,\\
[ {\mathcal{J}}^j_m, G_r^{ {\mathcal{S}} }   ]  
= & \frac{1}{2} \, \sigma_{ {\mathcal{S}}  {\mathcal{S}'} }^j  
\,  G_{m+r}^{ \mathcal{S}' } \, , \quad
[{\mathcal{J}}^j_m, H_r^{\mathcal{S}}   ]  
=  \frac{1}{2} \, \sigma_{ {\mathcal{S}} {\mathcal{S}}' }^j  
\,  H_{m+r}^{ {\mathcal{S}}' } \, , \\
[{\mathcal{J}}^j_m, {\mathcal{J}}^k_n   ] 
= &  i \, \epsilon_{ j k l}   \, {\mathcal{J}}^l_{m+n}\,,
\quad [ L_m, {\mathcal{J}}^j_n ] = - n \, {\mathcal{J}}^j_{ m+ n} \,,
\end{split}
\end{equation}
where $ ( {\mathcal{S}} , {\mathcal{S}}' )  = \pm $ , and $ {\sigma} ^j $ are the Pauli matrices. But we immediately see that the first two anti-commutators are incompatible, and hence this algebra is inadmissible. Similarly, an $SU(2)$ to rotate the $H_r^{\mathcal{S}} $'s only amongst themselves will also fail.

Next we can try to see if we can promote the $U(1)$ current algebra generated by $ S_n$'s to an $ SU(2)$ current algebra generated by $ {\mathcal{J}}^j_n$'s with the following modifications:
\begin{equation}
\begin{split}
\label{modi2}
 \left\{G_r^+, H_s^-\right\} = & M_{r+s} + \frac{1}{2}(r-s) \, \sigma_{+-}^j \, {\mathcal{J}}^j_{r+s}  \\
& + 2  \, C_2\left(r^2 - \frac{1}{4}\right)  \, \delta_{r+s,0}\, ,\\
 \left\{G_r^-, H_s^+\right\} = & M_{r+s} + \frac{1}{2}(r-s) \, \sigma_{-+}^j \, {\mathcal{J}}^j_{r+s}  \\
& + 2  \, C_2\left(r^2 - \frac{1}{4}\right)  \, \delta_{r+s,0}\, ,\\
\left\{ G_r^{ \mathcal{S} }, H_s^{ \mathcal{S}' } \right\} =
&  \frac{r-s}{2}  \,  
\sigma^3_{  \mathcal{S}   \mathcal{S}'  }  \, {\mathcal{J}}^3_{r+s}\,  ,\\
[ {\mathcal{J}}^j_m, G_r^{ + }   ]  
= & \frac{ \sigma_{ + - }^j }{2}  
\,  H_{m+r}^{- } +    \frac{ \sigma_{ + + }^j }{2}  
\,  G_{m+r}^{+ } \, , \\
[ {\mathcal{J}}^j_m, G_r^{ - }   ]  
= & \frac{ \sigma_{-+ }^j }{2}  
\,  H_{m+r}^{ + } +    \frac{ \sigma_{ -- }^j }{2}  
\,  G_{m+r}^{ - } \, , \\
[ {\mathcal{J}}^j_m, H_r^{ + }   ]  
= & \frac{ \sigma_{ + - }^j }{2}  
\,  G_{m+r}^{- } +    \frac{ \sigma_{ + + }^j }{2}  
\,  H_{m+r}^{+ } \, , \\
[ {\mathcal{J}}^j_m, H_r^{ - }   ]  
= & \frac{ \sigma_{ - + }^j }{2}  
\,  G_{m+r}^{ + } +    \frac{ \sigma_{ -- }^j }{2}  
\,  H_{m+r}^{- } \, , \\
[{\mathcal{J}}^j_m, {\mathcal{J}}^k_n   ] 
= &  i \, \epsilon_{j k l}   \, {\mathcal{J}}^l_{m+n}\,,
\quad [ L_m, {\mathcal{J}}^j_n ] = - n \, {\mathcal{J}}^j_{ m+ n} \,,
\end{split}
\end{equation}
where $ ( {\mathcal{S}} , {\mathcal{S}}' )  = \pm $. Here one can check that Jacobi identities
are not satisfied, for example the one involving $ \left ( \mathcal{J}_0^3 , G^+ _{\frac{1}{2}} , H^- _{ -\frac{1}{2}} 
\right )$. Hence this algebra is also inadmissible.

From the analysis above, we conclude that an $SU(2)$ extension of the R-symmetry does not seem feasible.

\section{Summary and Discussions}
\label{summary}

In this work, we have considered the SGCA in $2d$ with extended supercharges by taking a scaling limit (or group contraction) of the combination of the holomorphic and anti-holomorphic sectors of the $ {\mathcal{N}} = 2 $ $2d$ SCFT. This leads to the emergence of extra bosonic generators compared to the SGCA obtained from $\mathcal{N} = 1 $ SCFT, which are the analogues of the R-symmetry generator of the relativistic case. Assuming the state-operator correspondence, we have defined primary and descendent fields of this algebra and have derived their transformation rules under the action of the generators. We have also provided a superspace formalism in analogy with that of the ${ \mathcal{N} } = (2,2)$ SCFT. This allowed us to write the correlation functions of the superfields, which encode the correlators of the component fields. Lastly, we have proved that the $U(1) \times U (1)$ R-symmetry of the SGCA cannot be extended.

\section{Acknowledgements}

We thank Rajesh Gopakumar and Roji Pius for providing valuable inputs in improving the manuscript. I.M. was supported by NSERC of Canada and the Templeton
Foundation. A.R. is grateful for the hospitality provided by the Perimeter Institute during the Undergraduate Summer Research Program. Research at the Perimeter Institute is supported, in
part, by the Government of Canada through Industry Canada
and by the Province of Ontario through the Ministry of
Research and Information.

\clearpage

\bibliography{plb}

\begin{thebibliography}{47}%
\makeatletter
\providecommand \@ifxundefined [1]{%
 \@ifx{#1\undefined}
}%
\providecommand \@ifnum [1]{%
 \ifnum #1\expandafter \@firstoftwo
 \else \expandafter \@secondoftwo
 \fi
}%
\providecommand \@ifx [1]{%
 \ifx #1\expandafter \@firstoftwo
 \else \expandafter \@secondoftwo
 \fi
}%
\providecommand \natexlab [1]{#1}%
\providecommand \enquote  [1]{``#1''}%
\providecommand \bibnamefont  [1]{#1}%
\providecommand \bibfnamefont [1]{#1}%
\providecommand \citenamefont [1]{#1}%
\providecommand \href@noop [0]{\@secondoftwo}%
\providecommand \href [0]{\begingroup \@sanitize@url \@href}%
\providecommand \@href[1]{\@@startlink{#1}\@@href}%
\providecommand \@@href[1]{\endgroup#1\@@endlink}%
\providecommand \@sanitize@url [0]{\catcode `\\12\catcode `\$12\catcode
  `\&12\catcode `\#12\catcode `\^12\catcode `\_12\catcode `\%12\relax}%
\providecommand \@@startlink[1]{}%
\providecommand \@@endlink[0]{}%
\providecommand \url  [0]{\begingroup\@sanitize@url \@url }%
\providecommand \@url [1]{\endgroup\@href {#1}{\urlprefix }}%
\providecommand \urlprefix  [0]{URL }%
\providecommand \Eprint [0]{\href }%
\providecommand \doibase [0]{http://dx.doi.org/}%
\providecommand \selectlanguage [0]{\@gobble}%
\providecommand \bibinfo  [0]{\@secondoftwo}%
\providecommand \bibfield  [0]{\@secondoftwo}%
\providecommand \translation [1]{[#1]}%
\providecommand \BibitemOpen [0]{}%
\providecommand \bibitemStop [0]{}%
\providecommand \bibitemNoStop [0]{.\EOS\space}%
\providecommand \EOS [0]{\spacefactor3000\relax}%
\providecommand \BibitemShut  [1]{\csname bibitem#1\endcsname}%
\let\auto@bib@innerbib\@empty
\bibitem [{\citenamefont {{Bagchi}}\ and\ \citenamefont
  {{Gopakumar}}(2009)}]{gca}%
  \BibitemOpen
  \bibfield  {author} {\bibinfo {author} {\bibfnamefont {A.}~\bibnamefont
  {{Bagchi}}}\ and\ \bibinfo {author} {\bibfnamefont {R.}~\bibnamefont
  {{Gopakumar}}},\ }\href {\doibase 10.1088/1126-6708/2009/07/037} {\bibfield
  {journal} {\bibinfo  {journal} {Journal of High Energy Physics}\ }\textbf
  {\bibinfo {volume} {7}},\ \bibinfo {eid} {037} (\bibinfo {year} {2009})},\
  \Eprint {http://arxiv.org/abs/0902.1385} {arXiv:0902.1385 [hep-th]}
  \BibitemShut {NoStop}%
\bibitem [{\citenamefont {{Bagchi}}\ and\ \citenamefont
  {{Mandal}}(2009{\natexlab{a}})}]{ips2009}%
  \BibitemOpen
  \bibfield  {author} {\bibinfo {author} {\bibfnamefont {A.}~\bibnamefont
  {{Bagchi}}}\ and\ \bibinfo {author} {\bibfnamefont {I.}~\bibnamefont
  {{Mandal}}},\ }\href {\doibase 10.1016/j.physletb.2009.04.030} {\bibfield
  {journal} {\bibinfo  {journal} {Physics Letters B}\ }\textbf {\bibinfo
  {volume} {675}},\ \bibinfo {pages} {393} (\bibinfo {year}
  {2009}{\natexlab{a}})},\ \Eprint {http://arxiv.org/abs/0903.4524}
  {arXiv:0903.4524 [hep-th]} \BibitemShut {NoStop}%
\bibitem [{\citenamefont {{Bagchi}}\ \emph {et~al.}(2010)\citenamefont
  {{Bagchi}}, \citenamefont {{Gopakumar}}, \citenamefont {{Mandal}},\ and\
  \citenamefont {{Miwa}}}]{ba10}%
  \BibitemOpen
  \bibfield  {author} {\bibinfo {author} {\bibfnamefont {A.}~\bibnamefont
  {{Bagchi}}}, \bibinfo {author} {\bibfnamefont {R.}~\bibnamefont
  {{Gopakumar}}}, \bibinfo {author} {\bibfnamefont {I.}~\bibnamefont
  {{Mandal}}}, \ and\ \bibinfo {author} {\bibfnamefont {A.}~\bibnamefont
  {{Miwa}}},\ }\href {\doibase 10.1007/JHEP08(2010)004} {\bibfield  {journal}
  {\bibinfo  {journal} {Journal of High Energy Physics}\ }\textbf {\bibinfo
  {volume} {8}},\ \bibinfo {eid} {4} (\bibinfo {year} {2010})},\ \Eprint
  {http://arxiv.org/abs/0912.1090} {arXiv:0912.1090 [hep-th]} \BibitemShut
  {NoStop}%
\bibitem [{\citenamefont {{Bagchi}}\ and\ \citenamefont
  {{Kundu}}(2011)}]{bagchi11}%
  \BibitemOpen
  \bibfield  {author} {\bibinfo {author} {\bibfnamefont {A.}~\bibnamefont
  {{Bagchi}}}\ and\ \bibinfo {author} {\bibfnamefont {A.}~\bibnamefont
  {{Kundu}}},\ }\href {\doibase 10.1103/PhysRevD.83.066018} {\bibfield
  {journal} {\bibinfo  {journal} {Phys. Rev. D}\ }\textbf {\bibinfo {volume}
  {83}},\ \bibinfo {eid} {066018} (\bibinfo {year} {2011})},\ \Eprint
  {http://arxiv.org/abs/1011.4999} {arXiv:1011.4999 [hep-th]} \BibitemShut
  {NoStop}%
\bibitem [{\citenamefont {{Bagchi}}(2013)}]{bagchi13}%
  \BibitemOpen
  \bibfield  {author} {\bibinfo {author} {\bibfnamefont {A.}~\bibnamefont
  {{Bagchi}}},\ }\href {\doibase 10.1007/JHEP05(2013)141} {\bibfield  {journal}
  {\bibinfo  {journal} {Journal of High Energy Physics}\ }\textbf {\bibinfo
  {volume} {5}},\ \bibinfo {eid} {141} (\bibinfo {year} {2013})},\ \Eprint
  {http://arxiv.org/abs/1303.0291} {arXiv:1303.0291 [hep-th]} \BibitemShut
  {NoStop}%
\bibitem [{\citenamefont {{Hagen}}(1972)}]{hagen}%
  \BibitemOpen
  \bibfield  {author} {\bibinfo {author} {\bibfnamefont {C.~R.}\ \bibnamefont
  {{Hagen}}},\ }\href {\doibase 10.1103/PhysRevD.5.377} {\bibfield  {journal}
  {\bibinfo  {journal} {Phys. Rev. D}\ }\textbf {\bibinfo {volume} {5}},\
  \bibinfo {pages} {377} (\bibinfo {year} {1972})}\BibitemShut {NoStop}%
\bibitem [{\citenamefont {{Barnich}}\ \emph {et~al.}(2014)\citenamefont
  {{Barnich}}, \citenamefont {{Donnay}}, \citenamefont {{Matulich}},\ and\
  \citenamefont {{Troncoso}}}]{barnich}%
  \BibitemOpen
  \bibfield  {author} {\bibinfo {author} {\bibfnamefont {G.}~\bibnamefont
  {{Barnich}}}, \bibinfo {author} {\bibfnamefont {L.}~\bibnamefont {{Donnay}}},
  \bibinfo {author} {\bibfnamefont {J.}~\bibnamefont {{Matulich}}}, \ and\
  \bibinfo {author} {\bibfnamefont {R.}~\bibnamefont {{Troncoso}}},\ }\href
  {\doibase 10.1007/JHEP08(2014)071} {\bibfield  {journal} {\bibinfo  {journal}
  {Journal of High Energy Physics}\ }\textbf {\bibinfo {volume} {8}},\ \bibinfo
  {eid} {71} (\bibinfo {year} {2014})},\ \Eprint
  {http://arxiv.org/abs/1407.4275} {arXiv:1407.4275 [hep-th]} \BibitemShut
  {NoStop}%
\bibitem [{\citenamefont {{Oblak}}(2015)}]{oblak}%
  \BibitemOpen
  \bibfield  {author} {\bibinfo {author} {\bibfnamefont {B.}~\bibnamefont
  {{Oblak}}},\ }\href {\doibase 10.1007/s00220-015-2408-7} {\bibfield
  {journal} {\bibinfo  {journal} {Communications in Mathematical Physics}\
  }\textbf {\bibinfo {volume} {340}},\ \bibinfo {pages} {413} (\bibinfo {year}
  {2015})},\ \Eprint {http://arxiv.org/abs/1502.03108} {arXiv:1502.03108
  [hep-th]} \BibitemShut {NoStop}%
\bibitem [{\citenamefont {{Lukierski}}\ \emph {et~al.}(2006)\citenamefont
  {{Lukierski}}, \citenamefont {{Stichel}},\ and\ \citenamefont
  {{Zakrzewski}}}]{luk}%
  \BibitemOpen
  \bibfield  {author} {\bibinfo {author} {\bibfnamefont {J.}~\bibnamefont
  {{Lukierski}}}, \bibinfo {author} {\bibfnamefont {P.~C.}\ \bibnamefont
  {{Stichel}}}, \ and\ \bibinfo {author} {\bibfnamefont {W.~J.}\ \bibnamefont
  {{Zakrzewski}}},\ }\href {\doibase 10.1016/j.physleta.2006.04.016} {\bibfield
   {journal} {\bibinfo  {journal} {Physics Letters A}\ }\textbf {\bibinfo
  {volume} {357}},\ \bibinfo {pages} {1} (\bibinfo {year} {2006})},\ \Eprint
  {http://arxiv.org/abs/hep-th/0511259} {hep-th/0511259} \BibitemShut {NoStop}%
\bibitem [{\citenamefont {{de Azc{\'a}rraga}}\ and\ \citenamefont
  {{Lukierski}}(2009)}]{luk2}%
  \BibitemOpen
  \bibfield  {author} {\bibinfo {author} {\bibfnamefont {J.~A.}\ \bibnamefont
  {{de Azc{\'a}rraga}}}\ and\ \bibinfo {author} {\bibfnamefont
  {J.}~\bibnamefont {{Lukierski}}},\ }\href {\doibase
  10.1016/j.physletb.2009.06.042} {\bibfield  {journal} {\bibinfo  {journal}
  {Physics Letters B}\ }\textbf {\bibinfo {volume} {678}},\ \bibinfo {pages}
  {411} (\bibinfo {year} {2009})},\ \Eprint {http://arxiv.org/abs/0905.0141}
  {arXiv:0905.0141 [math-ph]} \BibitemShut {NoStop}%
\bibitem [{\citenamefont {{Duval}}\ and\ \citenamefont
  {{Horv{\'a}thy}}(2009)}]{duval}%
  \BibitemOpen
  \bibfield  {author} {\bibinfo {author} {\bibfnamefont {C.}~\bibnamefont
  {{Duval}}}\ and\ \bibinfo {author} {\bibfnamefont {P.~A.}\ \bibnamefont
  {{Horv{\'a}thy}}},\ }\href {\doibase 10.1088/1751-8113/42/46/465206}
  {\bibfield  {journal} {\bibinfo  {journal} {Journal of Physics A Mathematical
  General}\ }\textbf {\bibinfo {volume} {42}},\ \bibinfo {eid} {465206}
  (\bibinfo {year} {2009})},\ \Eprint {http://arxiv.org/abs/0904.0531}
  {arXiv:0904.0531 [math-ph]} \BibitemShut {NoStop}%
\bibitem [{\citenamefont {{Mukhopadhyay}}(2010)}]{ayan}%
  \BibitemOpen
  \bibfield  {author} {\bibinfo {author} {\bibfnamefont {A.}~\bibnamefont
  {{Mukhopadhyay}}},\ }\href {\doibase 10.1007/JHEP01(2010)100} {\bibfield
  {journal} {\bibinfo  {journal} {Journal of High Energy Physics}\ }\textbf
  {\bibinfo {volume} {1}},\ \bibinfo {eid} {100} (\bibinfo {year} {2010})},\
  \Eprint {http://arxiv.org/abs/0908.0797} {arXiv:0908.0797 [hep-th]}
  \BibitemShut {NoStop}%
\bibitem [{\citenamefont {{Martelli}}\ and\ \citenamefont
  {{Tachikawa}}(2010)}]{mart}%
  \BibitemOpen
  \bibfield  {author} {\bibinfo {author} {\bibfnamefont {D.}~\bibnamefont
  {{Martelli}}}\ and\ \bibinfo {author} {\bibfnamefont {Y.}~\bibnamefont
  {{Tachikawa}}},\ }\href {\doibase 10.1007/JHEP05(2010)091} {\bibfield
  {journal} {\bibinfo  {journal} {Journal of High Energy Physics}\ }\textbf
  {\bibinfo {volume} {5}},\ \bibinfo {eid} {91} (\bibinfo {year} {2010})},\
  \Eprint {http://arxiv.org/abs/0903.5184} {arXiv:0903.5184 [hep-th]}
  \BibitemShut {NoStop}%
\bibitem [{\citenamefont {{Hosseiny}}\ and\ \citenamefont
  {{Rouhani}}(2010)}]{rouhani}%
  \BibitemOpen
  \bibfield  {author} {\bibinfo {author} {\bibfnamefont {A.}~\bibnamefont
  {{Hosseiny}}}\ and\ \bibinfo {author} {\bibfnamefont {S.}~\bibnamefont
  {{Rouhani}}},\ }\href {\doibase 10.1063/1.3371191} {\bibfield  {journal}
  {\bibinfo  {journal} {Journal of Mathematical Physics}\ }\textbf {\bibinfo
  {volume} {51}},\ \bibinfo {pages} {052307} (\bibinfo {year} {2010})},\
  \Eprint {http://arxiv.org/abs/0909.1203} {arXiv:0909.1203 [hep-th]}
  \BibitemShut {NoStop}%
\bibitem [{\citenamefont {{Alishahiha}}\ \emph {et~al.}(2009)\citenamefont
  {{Alishahiha}}, \citenamefont {{Davody}},\ and\ \citenamefont
  {{Vahedi}}}]{ali}%
  \BibitemOpen
  \bibfield  {author} {\bibinfo {author} {\bibfnamefont {M.}~\bibnamefont
  {{Alishahiha}}}, \bibinfo {author} {\bibfnamefont {A.}~\bibnamefont
  {{Davody}}}, \ and\ \bibinfo {author} {\bibfnamefont {A.}~\bibnamefont
  {{Vahedi}}},\ }\href {\doibase 10.1088/1126-6708/2009/08/022} {\bibfield
  {journal} {\bibinfo  {journal} {Journal of High Energy Physics}\ }\textbf
  {\bibinfo {volume} {8}},\ \bibinfo {eid} {022} (\bibinfo {year} {2009})},\
  \Eprint {http://arxiv.org/abs/0903.3953} {arXiv:0903.3953 [hep-th]}
  \BibitemShut {NoStop}%
\bibitem [{\citenamefont {{Bagchi}}\ \emph
  {et~al.}(2015{\natexlab{a}})\citenamefont {{Bagchi}}, \citenamefont
  {{Chakrabortty}},\ and\ \citenamefont {{Parekh}}}]{arjuntensionless}%
  \BibitemOpen
  \bibfield  {author} {\bibinfo {author} {\bibfnamefont {A.}~\bibnamefont
  {{Bagchi}}}, \bibinfo {author} {\bibfnamefont {S.}~\bibnamefont
  {{Chakrabortty}}}, \ and\ \bibinfo {author} {\bibfnamefont {P.}~\bibnamefont
  {{Parekh}}},\ }\href@noop {} {\bibfield  {journal} {\bibinfo  {journal}
  {ArXiv e-prints}\ } (\bibinfo {year} {2015}{\natexlab{a}})},\ \Eprint
  {http://arxiv.org/abs/1507.04361} {arXiv:1507.04361 [hep-th]} \BibitemShut
  {NoStop}%
\bibitem [{\citenamefont {{Bagchi}}\ \emph
  {et~al.}(2015{\natexlab{b}})\citenamefont {{Bagchi}}, \citenamefont {{Basu}},
  \citenamefont {{Kakkar}},\ and\ \citenamefont {{Mehra}}}]{arjunym}%
  \BibitemOpen
  \bibfield  {author} {\bibinfo {author} {\bibfnamefont {A.}~\bibnamefont
  {{Bagchi}}}, \bibinfo {author} {\bibfnamefont {R.}~\bibnamefont {{Basu}}},
  \bibinfo {author} {\bibfnamefont {A.}~\bibnamefont {{Kakkar}}}, \ and\
  \bibinfo {author} {\bibfnamefont {A.}~\bibnamefont {{Mehra}}},\ }\href@noop
  {} {\bibfield  {journal} {\bibinfo  {journal} {ArXiv e-prints}\ } (\bibinfo
  {year} {2015}{\natexlab{b}})},\ \Eprint {http://arxiv.org/abs/1512.08375}
  {arXiv:1512.08375 [hep-th]} \BibitemShut {NoStop}%
\bibitem [{\citenamefont {{Bagchi}}\ and\ \citenamefont
  {{Mandal}}(2009{\natexlab{b}})}]{sgca4d}%
  \BibitemOpen
  \bibfield  {author} {\bibinfo {author} {\bibfnamefont {A.}~\bibnamefont
  {{Bagchi}}}\ and\ \bibinfo {author} {\bibfnamefont {I.}~\bibnamefont
  {{Mandal}}},\ }\href {\doibase 10.1103/PhysRevD.80.086011} {\bibfield
  {journal} {\bibinfo  {journal} {Phys. Rev. D}\ }\textbf {\bibinfo {volume}
  {80}},\ \bibinfo {eid} {086011} (\bibinfo {year} {2009}{\natexlab{b}})},\
  \Eprint {http://arxiv.org/abs/0905.0580} {arXiv:0905.0580 [hep-th]}
  \BibitemShut {NoStop}%
\bibitem [{\citenamefont {{Mandal}}(2010)}]{ma10}%
  \BibitemOpen
  \bibfield  {author} {\bibinfo {author} {\bibfnamefont {I.}~\bibnamefont
  {{Mandal}}},\ }\href {\doibase 10.1007/JHEP11(2010)018} {\bibfield  {journal}
  {\bibinfo  {journal} {Journal of High Energy Physics}\ }\textbf {\bibinfo
  {volume} {11}},\ \bibinfo {eid} {18} (\bibinfo {year} {2010})},\ \Eprint
  {http://arxiv.org/abs/1003.0209} {arXiv:1003.0209 [hep-th]} \BibitemShut
  {NoStop}%
\bibitem [{\citenamefont {{Aizawa}}(2012)}]{aizawa0}%
  \BibitemOpen
  \bibfield  {author} {\bibinfo {author} {\bibfnamefont {N.}~\bibnamefont
  {{Aizawa}}},\ }\href {\doibase 10.1088/1751-8113/45/47/475203} {\bibfield
  {journal} {\bibinfo  {journal} {Journal of Physics A Mathematical General}\
  }\textbf {\bibinfo {volume} {45}},\ \bibinfo {eid} {475203} (\bibinfo {year}
  {2012})},\ \Eprint {http://arxiv.org/abs/1206.2708} {arXiv:1206.2708
  [math-ph]} \BibitemShut {NoStop}%
\bibitem [{\citenamefont {{Gao}}\ \emph {et~al.}(2014)\citenamefont {{Gao}},
  \citenamefont {{Pei}},\ and\ \citenamefont {{Bai}}}]{gao}%
  \BibitemOpen
  \bibfield  {author} {\bibinfo {author} {\bibfnamefont {S.}~\bibnamefont
  {{Gao}}}, \bibinfo {author} {\bibfnamefont {Y.}~\bibnamefont {{Pei}}}, \ and\
  \bibinfo {author} {\bibfnamefont {C.}~\bibnamefont {{Bai}}},\ }\href
  {\doibase 10.1088/1751-8113/47/22/225202} {\bibfield  {journal} {\bibinfo
  {journal} {Journal of Physics A Mathematical General}\ }\textbf {\bibinfo
  {volume} {47}},\ \bibinfo {eid} {225202} (\bibinfo {year}
  {2014})}\BibitemShut {NoStop}%
\bibitem [{\citenamefont {{Aizawa}}\ \emph {et~al.}(2013)\citenamefont
  {{Aizawa}}, \citenamefont {{Kuznetsova}},\ and\ \citenamefont
  {{Toppan}}}]{aizawa1}%
  \BibitemOpen
  \bibfield  {author} {\bibinfo {author} {\bibfnamefont {N.}~\bibnamefont
  {{Aizawa}}}, \bibinfo {author} {\bibfnamefont {Z.}~\bibnamefont
  {{Kuznetsova}}}, \ and\ \bibinfo {author} {\bibfnamefont {F.}~\bibnamefont
  {{Toppan}}},\ }\href {\doibase 10.1063/1.4820481} {\bibfield  {journal}
  {\bibinfo  {journal} {Journal of Mathematical Physics}\ }\textbf {\bibinfo
  {volume} {54}},\ \bibinfo {pages} {093506} (\bibinfo {year} {2013})},\
  \Eprint {http://arxiv.org/abs/1307.5259} {arXiv:1307.5259 [hep-th]}
  \BibitemShut {NoStop}%
\bibitem [{\citenamefont {{Masterov}}(2012)}]{masterov12}%
  \BibitemOpen
  \bibfield  {author} {\bibinfo {author} {\bibfnamefont {I.}~\bibnamefont
  {{Masterov}}},\ }\href {\doibase 10.1063/1.4732459} {\bibfield  {journal}
  {\bibinfo  {journal} {Journal of Mathematical Physics}\ }\textbf {\bibinfo
  {volume} {53}},\ \bibinfo {pages} {072904} (\bibinfo {year} {2012})},\
  \Eprint {http://arxiv.org/abs/1112.4924} {arXiv:1112.4924 [hep-th]}
  \BibitemShut {NoStop}%
\bibitem [{\citenamefont {{Masterov}}(2015)}]{masterov15}%
  \BibitemOpen
  \bibfield  {author} {\bibinfo {author} {\bibfnamefont {I.}~\bibnamefont
  {{Masterov}}},\ }\href {\doibase 10.1063/1.4909528} {\bibfield  {journal}
  {\bibinfo  {journal} {Journal of Mathematical Physics}\ }\textbf {\bibinfo
  {volume} {56}},\ \bibinfo {eid} {022902} (\bibinfo {year} {2015})},\ \Eprint
  {http://arxiv.org/abs/1410.5335} {arXiv:1410.5335 [hep-th]} \BibitemShut
  {NoStop}%
\bibitem [{\citenamefont {{Sakaguchi}}(2010)}]{sakaguchi}%
  \BibitemOpen
  \bibfield  {author} {\bibinfo {author} {\bibfnamefont {M.}~\bibnamefont
  {{Sakaguchi}}},\ }\href {\doibase 10.1063/1.3321531} {\bibfield  {journal}
  {\bibinfo  {journal} {Journal of Mathematical Physics}\ }\textbf {\bibinfo
  {volume} {51}},\ \bibinfo {pages} {042301} (\bibinfo {year} {2010})},\
  \Eprint {http://arxiv.org/abs/0905.0188} {arXiv:0905.0188 [hep-th]}
  \BibitemShut {NoStop}%
\bibitem [{\citenamefont {Boucher}\ \emph {et~al.}(1986)\citenamefont
  {Boucher}, \citenamefont {Friedan},\ and\ \citenamefont {Kent}}]{boucher}%
  \BibitemOpen
  \bibfield  {author} {\bibinfo {author} {\bibfnamefont {W.}~\bibnamefont
  {Boucher}}, \bibinfo {author} {\bibfnamefont {D.}~\bibnamefont {Friedan}}, \
  and\ \bibinfo {author} {\bibfnamefont {A.}~\bibnamefont {Kent}},\ }\href@noop
  {} {\bibfield  {journal} {\bibinfo  {journal} {Physics Letters B}\ }\textbf
  {\bibinfo {volume} {172}},\ \bibinfo {pages} {316} (\bibinfo {year}
  {1986})}\BibitemShut {NoStop}%
\bibitem [{\citenamefont {Sen}(1987)}]{sen}%
  \BibitemOpen
  \bibfield  {author} {\bibinfo {author} {\bibfnamefont {A.}~\bibnamefont
  {Sen}},\ }\href {\doibase http://dx.doi.org/10.1016/0550-3213(87)90043-5}
  {\bibfield  {journal} {\bibinfo  {journal} {Nuclear Physics B}\ }\textbf
  {\bibinfo {volume} {284}},\ \bibinfo {pages} {423 } (\bibinfo {year}
  {1987})}\BibitemShut {NoStop}%
\bibitem [{\citenamefont {Banks}\ \emph {et~al.}(1988)\citenamefont {Banks},
  \citenamefont {Dixon}, \citenamefont {Friedan},\ and\ \citenamefont
  {Martinec}}]{banks}%
  \BibitemOpen
  \bibfield  {author} {\bibinfo {author} {\bibfnamefont {T.}~\bibnamefont
  {Banks}}, \bibinfo {author} {\bibfnamefont {L.~J.}\ \bibnamefont {Dixon}},
  \bibinfo {author} {\bibfnamefont {D.}~\bibnamefont {Friedan}}, \ and\
  \bibinfo {author} {\bibfnamefont {E.}~\bibnamefont {Martinec}},\ }\href
  {\doibase http://dx.doi.org/10.1016/0550-3213(88)90551-2} {\bibfield
  {journal} {\bibinfo  {journal} {Nuclear Physics B}\ }\textbf {\bibinfo
  {volume} {299}},\ \bibinfo {pages} {613 } (\bibinfo {year}
  {1988})}\BibitemShut {NoStop}%
\bibitem [{\citenamefont {{Friedan}}\ \emph {et~al.}(1984)\citenamefont
  {{Friedan}}, \citenamefont {{Qiu}},\ and\ \citenamefont {{Shenker}}}]{qiu84}%
  \BibitemOpen
  \bibfield  {author} {\bibinfo {author} {\bibfnamefont {D.}~\bibnamefont
  {{Friedan}}}, \bibinfo {author} {\bibfnamefont {Z.}~\bibnamefont {{Qiu}}}, \
  and\ \bibinfo {author} {\bibfnamefont {S.}~\bibnamefont {{Shenker}}},\ }\href
  {\doibase 10.1103/PhysRevLett.52.1575} {\bibfield  {journal} {\bibinfo
  {journal} {Physical Review Letters}\ }\textbf {\bibinfo {volume} {52}},\
  \bibinfo {pages} {1575} (\bibinfo {year} {1984})}\BibitemShut {NoStop}%
\bibitem [{\citenamefont {{Friedan}}\ \emph {et~al.}(1985)\citenamefont
  {{Friedan}}, \citenamefont {{Qiu}},\ and\ \citenamefont {{Shenker}}}]{qiu85}%
  \BibitemOpen
  \bibfield  {author} {\bibinfo {author} {\bibfnamefont {D.}~\bibnamefont
  {{Friedan}}}, \bibinfo {author} {\bibfnamefont {Z.}~\bibnamefont {{Qiu}}}, \
  and\ \bibinfo {author} {\bibfnamefont {S.}~\bibnamefont {{Shenker}}},\ }\href
  {\doibase 10.1016/0370-2693(85)90819-6} {\bibfield  {journal} {\bibinfo
  {journal} {Physics Letters B}\ }\textbf {\bibinfo {volume} {151}},\ \bibinfo
  {pages} {37} (\bibinfo {year} {1985})}\BibitemShut {NoStop}%
\bibitem [{\citenamefont {{Qiu}}(1986)}]{qiu86}%
  \BibitemOpen
  \bibfield  {author} {\bibinfo {author} {\bibfnamefont {Z.}~\bibnamefont
  {{Qiu}}},\ }\href {\doibase 10.1016/0550-3213(86)90553-5} {\bibfield
  {journal} {\bibinfo  {journal} {Nuclear Physics B}\ }\textbf {\bibinfo
  {volume} {270}},\ \bibinfo {pages} {205} (\bibinfo {year}
  {1986})}\BibitemShut {NoStop}%
\bibitem [{\citenamefont {{Bershadsky}}\ \emph {et~al.}(1985)\citenamefont
  {{Bershadsky}}, \citenamefont {{Knizhnik}},\ and\ \citenamefont
  {{Teitelman}}}]{berhadsky}%
  \BibitemOpen
  \bibfield  {author} {\bibinfo {author} {\bibfnamefont {M.~A.}\ \bibnamefont
  {{Bershadsky}}}, \bibinfo {author} {\bibfnamefont {V.~G.}\ \bibnamefont
  {{Knizhnik}}}, \ and\ \bibinfo {author} {\bibfnamefont {M.~G.}\ \bibnamefont
  {{Teitelman}}},\ }\href {\doibase 10.1016/0370-2693(85)90818-4} {\bibfield
  {journal} {\bibinfo  {journal} {Physics Letters B}\ }\textbf {\bibinfo
  {volume} {151}},\ \bibinfo {pages} {31} (\bibinfo {year} {1985})}\BibitemShut
  {NoStop}%
\bibitem [{\citenamefont {{Goddard}}\ \emph {et~al.}(1986)\citenamefont
  {{Goddard}}, \citenamefont {{Kent}},\ and\ \citenamefont
  {{Olive}}}]{goddard}%
  \BibitemOpen
  \bibfield  {author} {\bibinfo {author} {\bibfnamefont {P.}~\bibnamefont
  {{Goddard}}}, \bibinfo {author} {\bibfnamefont {A.}~\bibnamefont {{Kent}}}, \
  and\ \bibinfo {author} {\bibfnamefont {D.}~\bibnamefont {{Olive}}},\ }\href
  {\doibase 10.1007/BF01464283} {\bibfield  {journal} {\bibinfo  {journal}
  {Communications in Mathematical Physics}\ }\textbf {\bibinfo {volume}
  {103}},\ \bibinfo {pages} {105} (\bibinfo {year} {1986})}\BibitemShut
  {NoStop}%
\bibitem [{\citenamefont {{Sotkov}}\ and\ \citenamefont
  {{Stanishkov}}(1986)}]{sotkov}%
  \BibitemOpen
  \bibfield  {author} {\bibinfo {author} {\bibfnamefont {G.~M.}\ \bibnamefont
  {{Sotkov}}}\ and\ \bibinfo {author} {\bibfnamefont {M.~S.}\ \bibnamefont
  {{Stanishkov}}},\ }\href {\doibase 10.1016/0370-2693(86)90768-9} {\bibfield
  {journal} {\bibinfo  {journal} {Physics Letters B}\ }\textbf {\bibinfo
  {volume} {177}},\ \bibinfo {pages} {361} (\bibinfo {year}
  {1986})}\BibitemShut {NoStop}%
\bibitem [{\citenamefont {Zamolodchikov}\ and\ \citenamefont
  {Fateev}(1986)}]{Zamolodchikov86}%
  \BibitemOpen
  \bibfield  {author} {\bibinfo {author} {\bibfnamefont {A.~B.}\ \bibnamefont
  {Zamolodchikov}}\ and\ \bibinfo {author} {\bibfnamefont {V.~A.}\ \bibnamefont
  {Fateev}},\ }\href@noop {} {\bibfield  {journal} {\bibinfo  {journal} {Sov.
  Phys. JETP}\ }\textbf {\bibinfo {volume} {63}},\ \bibinfo {pages} {913}
  (\bibinfo {year} {1986})},\ \bibinfo {note} {[Zh. Eksp. Teor.
  Fiz.90,1553(1986)]}\BibitemShut {NoStop}%
\bibitem [{\citenamefont {Di~Vecchia}\ \emph {et~al.}(1985)\citenamefont
  {Di~Vecchia}, \citenamefont {Petersen},\ and\ \citenamefont {Zheng}}]{di85}%
  \BibitemOpen
  \bibfield  {author} {\bibinfo {author} {\bibfnamefont {P.}~\bibnamefont
  {Di~Vecchia}}, \bibinfo {author} {\bibfnamefont {J.}~\bibnamefont
  {Petersen}}, \ and\ \bibinfo {author} {\bibfnamefont {H.}~\bibnamefont
  {Zheng}},\ }\href@noop {} {\bibfield  {journal} {\bibinfo  {journal} {Physics
  Letters B}\ }\textbf {\bibinfo {volume} {162}},\ \bibinfo {pages} {327}
  (\bibinfo {year} {1985})}\BibitemShut {NoStop}%
\bibitem [{\citenamefont {Di~Vecchia}\ \emph
  {et~al.}(1986{\natexlab{a}})\citenamefont {Di~Vecchia}, \citenamefont
  {Petersen},\ and\ \citenamefont {Yu}}]{di86}%
  \BibitemOpen
  \bibfield  {author} {\bibinfo {author} {\bibfnamefont {P.}~\bibnamefont
  {Di~Vecchia}}, \bibinfo {author} {\bibfnamefont {J.}~\bibnamefont
  {Petersen}}, \ and\ \bibinfo {author} {\bibfnamefont {M.}~\bibnamefont
  {Yu}},\ }\href@noop {} {\bibfield  {journal} {\bibinfo  {journal} {Physics
  Letters B}\ }\textbf {\bibinfo {volume} {172}},\ \bibinfo {pages} {211}
  (\bibinfo {year} {1986}{\natexlab{a}})}\BibitemShut {NoStop}%
\bibitem [{\citenamefont {Di~Vecchia}\ \emph
  {et~al.}(1986{\natexlab{b}})\citenamefont {Di~Vecchia}, \citenamefont
  {Petersen}, \citenamefont {Yu},\ and\ \citenamefont {Zheng}}]{di86-2}%
  \BibitemOpen
  \bibfield  {author} {\bibinfo {author} {\bibfnamefont {P.}~\bibnamefont
  {Di~Vecchia}}, \bibinfo {author} {\bibfnamefont {J.}~\bibnamefont
  {Petersen}}, \bibinfo {author} {\bibfnamefont {M.}~\bibnamefont {Yu}}, \ and\
  \bibinfo {author} {\bibfnamefont {H.}~\bibnamefont {Zheng}},\ }\href@noop {}
  {\bibfield  {journal} {\bibinfo  {journal} {Physics Letters B}\ }\textbf
  {\bibinfo {volume} {174}},\ \bibinfo {pages} {280} (\bibinfo {year}
  {1986}{\natexlab{b}})}\BibitemShut {NoStop}%
\bibitem [{\citenamefont {Nam}(1986)}]{nam}%
  \BibitemOpen
  \bibfield  {author} {\bibinfo {author} {\bibfnamefont {S.}~\bibnamefont
  {Nam}},\ }\href@noop {} {\bibfield  {journal} {\bibinfo  {journal} {Physics
  Letters B}\ }\textbf {\bibinfo {volume} {172}},\ \bibinfo {pages} {323}
  (\bibinfo {year} {1986})}\BibitemShut {NoStop}%
\bibitem [{\citenamefont {Dobrev}(1987)}]{dobrev}%
  \BibitemOpen
  \bibfield  {author} {\bibinfo {author} {\bibfnamefont {V.}~\bibnamefont
  {Dobrev}},\ }\href@noop {} {\bibfield  {journal} {\bibinfo  {journal}
  {Physics Letters B}\ }\textbf {\bibinfo {volume} {186}},\ \bibinfo {pages}
  {43} (\bibinfo {year} {1987})}\BibitemShut {NoStop}%
\bibitem [{\citenamefont {Matsuo}(1987)}]{matsuo}%
  \BibitemOpen
  \bibfield  {author} {\bibinfo {author} {\bibfnamefont {Y.}~\bibnamefont
  {Matsuo}},\ }\href@noop {} {\bibfield  {journal} {\bibinfo  {journal}
  {Progress of theoretical physics}\ }\textbf {\bibinfo {volume} {77}},\
  \bibinfo {pages} {793} (\bibinfo {year} {1987})}\BibitemShut {NoStop}%
\bibitem [{\citenamefont {Kiritsis}(1987)}]{kiritsis}%
  \BibitemOpen
  \bibfield  {author} {\bibinfo {author} {\bibfnamefont {E.~B.}\ \bibnamefont
  {Kiritsis}},\ }\href {\doibase 10.1103/PhysRevD.36.3048} {\bibfield
  {journal} {\bibinfo  {journal} {Phys. Rev. D}\ }\textbf {\bibinfo {volume}
  {36}},\ \bibinfo {pages} {3048} (\bibinfo {year} {1987})}\BibitemShut
  {NoStop}%
\bibitem [{\citenamefont {{Gieres}}\ and\ \citenamefont
  {{Gourmelen}}(1997)}]{gi97}%
  \BibitemOpen
  \bibfield  {author} {\bibinfo {author} {\bibfnamefont {F.}~\bibnamefont
  {{Gieres}}}\ and\ \bibinfo {author} {\bibfnamefont {S.}~\bibnamefont
  {{Gourmelen}}},\ }in\ \href@noop {} {\emph {\bibinfo {booktitle} {eprint
  arXiv:solv-int/9708009}}}\ (\bibinfo {year} {1997})\ p.\ \bibinfo {pages}
  {8009}\BibitemShut {NoStop}%
\bibitem [{\citenamefont {West}(1990)}]{we90}%
  \BibitemOpen
  \bibfield  {author} {\bibinfo {author} {\bibfnamefont {P.}~\bibnamefont
  {West}},\ }\href@noop {} {\emph {\bibinfo {title} {Introduction to
  Supersymmetry and Supergravity}}}\ (\bibinfo  {publisher} {World
  Scientific},\ \bibinfo {year} {1990})\BibitemShut {NoStop}%
\bibitem [{\citenamefont {Blumenhagen}\ and\ \citenamefont
  {Plauschinn}(2009)}]{blu09}%
  \BibitemOpen
  \bibfield  {author} {\bibinfo {author} {\bibfnamefont {R.}~\bibnamefont
  {Blumenhagen}}\ and\ \bibinfo {author} {\bibfnamefont {E.}~\bibnamefont
  {Plauschinn}},\ }\href@noop {} {\emph {\bibinfo {title} {Introduction to
  Conformal Field Theory: With Applications to String Theory}}},\ Lecture Notes
  in Physics\ (\bibinfo  {publisher} {Springer},\ \bibinfo {year}
  {2009})\BibitemShut {NoStop}%
\bibitem [{\citenamefont {{Schwimmer}}\ and\ \citenamefont
  {{Seiberg}}(1987)}]{seiberg}%
  \BibitemOpen
  \bibfield  {author} {\bibinfo {author} {\bibfnamefont {A.}~\bibnamefont
  {{Schwimmer}}}\ and\ \bibinfo {author} {\bibfnamefont {N.}~\bibnamefont
  {{Seiberg}}},\ }\href {\doibase 10.1016/0370-2693(87)90566-1} {\bibfield
  {journal} {\bibinfo  {journal} {Physics Letters B}\ }\textbf {\bibinfo
  {volume} {184}},\ \bibinfo {pages} {191} (\bibinfo {year}
  {1987})}\BibitemShut {NoStop}%
\bibitem [{\citenamefont {{Inonu}}\ and\ \citenamefont
  {{Wigner}}(1953)}]{inonu}%
  \BibitemOpen
  \bibfield  {author} {\bibinfo {author} {\bibfnamefont {E.}~\bibnamefont
  {{Inonu}}}\ and\ \bibinfo {author} {\bibfnamefont {E.~P.}\ \bibnamefont
  {{Wigner}}},\ }\href {\doibase 10.1073/pnas.39.6.510} {\bibfield  {journal}
  {\bibinfo  {journal} {Proceedings of the National Academy of Science}\
  }\textbf {\bibinfo {volume} {39}},\ \bibinfo {pages} {510} (\bibinfo {year}
  {1953})}\BibitemShut {NoStop}%
\end{thebibliography}%


\end{document}